# Starburst99 for Windows

Claus Leitherer and Julia Chen

*Space Telescope Science Institute[1], 3700 San Martin Drive, Baltimore, MD 21218*

leitherer@stsci.edu, zchen@stsci.edu

## Abstract

We describe a Windows compatible version of the evolutionary synthesis code Starburst99. Starburst99 for Windows was developed from the public UNIX based version at STScI. We converted the original Fortran77 source code into a version for a Win32 environment with an Absoft[2] Fortran Pro x86 compiler. Extensive testing showed no significant numerical differences in comparison with the previous UNIX version. The software application consists of the source code, executable, and a number of auxiliary files. The package installs on any PC running Windows 2000, XP, or Vista and can be obtained as freeware at http://www.stsci.edu/science/starburst/PCStarburst99.html. We give an overview of the different running modes and provide instructions for getting started with the initial set-up.

*Key words*: Data Analysis and Techniques

---

# 1. Introduction

Starburst99 is a widely used software package for evolutionary synthesis of young stellar populations. Its goal is to support astronomers in their interpretation of spectra, spectral energy distributions and other observational data of star clusters and galaxies. The package evolved from a simple, Fortran77-based software tool to calculate basic stellar properties, such as luminosities, ionizing photon output, and number distributions of stellar types (Leitherer 1990). Over time, additional capabilities such as supernova rates and wind properties were added (Leitherer, Robert, & Drissen 1992). Subsequently, spectral synthesis of line profiles was implemented (Robert, Leitherer, & Heckman 1993; 1995). Leitherer & Heckman (1995) provided a compilation of the capabilities of the available software, including a detailed parameter study. That work was performed at the dawn of the internet age. A fixed grid of standard models was calculated and made available on our website.

The term "Starburst99" was coined during a summer student project when D. Foo Kune ported the software to the Space Telescope Science Institute (STScI) web server in 1998 (Leitherer et al. 1999; hereafter L99). The package was made publicly available for download, and interested users could remotely run model simulations on the STScI server at http://www.stsci.edu/science/starburst99/. As of 2008, this set-up is still in place with few technical changes. The impact of Starburst99 can be gauged from the more than 1000 citations to L99 (as of 2008), which is the most commonly used reference to the Starburst99 software. However, major upgrades to the software have been performed since the original release. Additional spectral synthesis capabilities were introduced in the infrared (Origlia et al. 1999), ultraviolet (González Delgado, Leitherer, & Heckman 1997; de Mello, Leitherer, & Heckman 2000; Leitherer et al. 2001; Robert et al. 2003), and optical (González Delgado & Leitherer 1999; González Delgado et al. 1999; González Delgado et al. 2005; Martins et al. 2005).

The currently running version 5.1 of Starburst99 was described by Vázquez & Leitherer (2005), who expanded the choices of stellar evolution models to allow modeling of young as well as old stellar populations. The reader is referred to that paper and to L99 for a discussion of the input physics and the astronomical aspects of Starburst99 since the main focus of the present paper is on software and implementation issues.

The Starburst99 source code was originally written in VMS Fortran77 and converted to UNIX/Solaris prior to release on our web server in 1998. Currently the software is running under Solaris v10. During the past decade the use of personal computers has seen a steady increase at the expense of Sun workstations. In parallel with this shift towards PCs, computers running UNIX software have become less prevalent (at least in research environments). As a result, there was demand for a version of the Starburst99 software package that would run on a PC. A LINUX version of the source code was generated and distributed in 2004.[3]

In this paper we describe a version of the Starburst99 software that can be installed, re-compiled (if desired), and run on Windows based PCs. Our effort was motivated by the flexibility offered by a user-owned desktop or notebook, as opposed to the limitations of the STScI web server where users have to compete with others for system resources. There are obviously additional advantages: users can modify the source code according to their needs, there is no dependence on network connectivity (which can be a challenge for GB-sized outputs), and the models can immediately be visualized and compared to data. In Section 2 of this paper we describe the source code and the structure of the file system. The conversion from UNIX to Windows Fortran is discussed in Section 3. The installation of the software package is covered in Section 4. Benchmark tests and a performance description are in Section 5. In Section 6 we provide instructions on how to obtain, install, and run the software on a PC.

## 2. Source Code and File Structure

The source code for Starburst99 was originally written in Fortran77 and ran in a VMS environment on a mainframe computer until the mid-1990's. When computers running the VMS operating system became largely obsolete, the code was ported to a UNIX/Solaris environment. We briefly experimented with converting from Fortran77 to C++. However it quickly became clear that the C++ version would not be very useful unless an enormous amount of effort and resources would be spent. The C++ code we generated reproduced the numerical results of the Fortran77 code but was slower by a factor of several and was much harder to understand and manipulate for many non-specialists who are used to Fortran. In the end we decided to stick with Fortran despite some arguments that can be made against this programming language (Ferland 2000).

---

[3] L. Yao (Univ. of Toronto) kindly prepared the LINUX version of Starburst99.

Although originally written in Fortran77, the source code now uses both Fortran77 and Fortran90 based syntax. This is the result of numerous contributors providing subroutines and updates to the code. The hybrid structure generally causes no problems, as Fortran90 is backward compatible with Fortran77.

The current version 5.1 of the Starburst99 source code has about 11,000 lines of code. It is structured into a main program, subroutines, functions, and a block data section. The only purpose of the main program is to perform a full duty cycle in time and to call each subroutine as needed. The majority of the subroutines are called only if requested by the user in order to economize system resources. Most variables and arrays are passed along via common blocks. The arrays have dimensions of 3 or less. The sizes of the largest arrays are determined by the stellar mass grid and the wavelength points of the stellar spectra, which can be as large as 2000 and 14000, respectively. There are no calls to external mathematical libraries. In order to maximize portability, mathematical utilities (such as interpolation and integration routines) are explicitly included as subroutines and functions.

Input and output are performed via text files that are read in and out by the software, respectively. Starburst99 does not produce any graphics. This capability may be added externally by the user. Both input and output are in text format and can be displayed with a standard editor. In particular, the output files are well suited for further manipulation or display with IDL software. The Starburst99 package includes a number of auxiliary files, all of which are in text format as well. These files contain supporting libraries and look-up tables for ingestion by the code. The full list and description of these files are given in Section 4.

The public UNIX/Solaris version of Starburst99 is installed on a Sun Blade-100 workstation and can be accessed at http://www.stsci.edu/science/starburst99/. This workstation has a 500 MHz UltraSPARC-IIe CPU and a physical memory of 1.75 GB. In the absence of any other load, running Starburst99 in its default parameter mode takes 4.5 hours to complete.

## 3. Windows Fortran Conversion

After our unsatisfactory attempts of converting the Fortran code into C++, which can directly be compiled on a Microsoft Windows platform, we decided to go a different route, i.e., utilizing a product that can compile UNIX-based Fortran code on Microsoft Windows. After performing an extensive trade study of several freeware as well as commercial products, we selected Absoft Pro Fortran 10.0.4 for Windows. The principal reasons were:

- The product meets the fundamental need of compiling UNIX-based Fortran code on the Windows platform.
- The license fee is reasonable and includes excellent user support. An adequate level of user support turned out to be essential during the course of this project.
- Once the source code is successfully compiled and linked, the output file can be executed on a Windows platform without any Absoft libraries.

The latter point was particular relevant for a decision in favor of the Absoft product. Starburst99 is used by a diverse user community with varying software resources and budgets. As a result, many users do not have a Windows Fortran compiler available and would not be able to use the distributed software package if the executable were dependent on the installed Absoft product. Since the compiled executable is fully self-contained and does not need any additional external libraries, it can easily be copied and transferred as desired. Of course, a user cannot make modifications to the distributed source code itself without having access to a suitable Fortran compiler.

The conversion of the original UNIX-based Fortran source code to a Windows compatible executable required rather extensive re-coding. Most of the changes were required because of the existing vector and array initialization, which was incompatible with Windows Fortran. While arrays are automatically reset to zero during the first call in a subroutine in UNIX, this is not the case in Windows. Therefore a large number of vectors and arrays had to be modified and restructured. As a benefit of this exercise, the revised code is now backward compatible with a UNIX environment.

The compile and link procedure was facilitated with the standard Absoft Fortran Pro interface. The distributed executable was generated with the following compiler options:

- *–stack:350000000*: This option allocates sufficient memory initially, rather than letting Windows increase it incrementally.
- *–f*: All symbols in the source code are folded to lower case.

- *–yext_names="lcs"*: This option specifies lower case for all external names as they appear to other object files.
- *–yext_sfx="_"*: The user specified string "_" is appended to the external representation of procedure names.
- *–mno-sse2*: The use of the SSE2[4] instruction for floating point operations is disabled. SSE2 was introduced by Intel with the Pentium 4 processor and would not work with older PCs.
- *–unix.lib, –vms.lib*: The library names are passed to the linker, ensuring that some of the legacy nomenclature is working.

## 4. Windows Installation

After the source code was successfully compiled and the executable generated, we moved on to our next goal, which was to provide an auto-install package that installs the product and also provides a user-friendly interface. For this task, we chose Microsoft Visual Studio 2005 for its rich and rapid Integrated Development Environment. As the existing web user interface for the Unix-based program had been widely used by the community, we chose to provide a similar Windows-based user interface for consistency.

The Windows Installer assigns a root folder to all Starburst99 files in the "Program Files" folder during the installation. The root folder contains, among other files, the source code and the executable. Subfolders contain the help files, the model output, and various data files in support of Starburst99. There are three groups of supporting data files. (i) 18 input files contain the stellar evolution models from both the Geneva and Padova groups. A summary of the file properties and the nomenclature is in Table 1. A detailed description of the four sets of evolutionary tracks is in L99 and Vázquez & Leitherer (2005). Each of the four sets is available for a range of heavy-element abundances. (ii) The theoretical model atmospheres are in 42 different input files. The files are listed in Table 2. As for the evolution models, the atmospheres are available for a set of abundances. Note that the files "ifa" are not currently used and are intended for future implementation. More information on theses files and on the physics can be found in the previously cited literature and in the Help files that come as part of the package. (iii) Six auxiliary files contain sets of empirical libraries and look-up tables with stellar properties (Table 3). While all files are in text format, they were not generated for easy reading by humans but for maximum compactness. Explanations are in the Help files and on the STScI Starburst99 web site.

The installation of the software follows the standard Windows procedure: icons and links are added to the desktop, taskbar and quick-launch bar. The application is launched by clicking on any of these icons. Upon initialization, the graphic Starburst99 interface opens. Our design of the interface is similar to the one on the Starburst99 web server, with some enhancements using richer functions offered by Windows. Some of the enhancements are as follows.

We grouped the input parameters into several tabs in order to avoid the need for scrolling. The first set of parameters is related to star formation (see Fig. 1). Users can adopt the suggested set of standard parameters or change the input according to their needs. All input must be formatted in English nomenclature and style, and the Windows Regional and Language settings must support English. In particular, decimal numbers must use periods, not commas. The other groups of input parameters are related to stellar evolution (Fig. 2) and stellar atmospheres and spectra (Fig. 3). When the range of input parameters is not freely selectable, users can select the parameters from pull-down menus. The desired output files can be selected individually (Fig. 4). Since some of the generated files can be rather large, user may prefer to unselect them if they are not needed. Note that "Ifaspec" will be a future implementation. Some output files are interrelated, i.e., one cannot be generated if another one is not available. The Help file explains these dependencies. In addition, a warning and/or error messages will be displayed in the Runtime Message window (Fig. 5). This window automatically opens when relevant information for the users becomes available.

---

[4] SSE2= Streaming SIMD Extensions2, where SIMD means single instruction, multiple data

The output files, including a log of the model run, can be accessed via a Windows-type graphical interface (Fig. 6). The interface allows both selection and viewing of the files. No graphical display software for the output files is provided. Users are expected to user their own software such as, e.g., IDL. Additional features of the user interface are a web link for product registration and notification of future releases and enhancements and access to extensive Help files.

## 5. Performance and Benchmark Tests

The Starburst99 application is configured to run with a set of default parameters which produce sensible results. A standard set of output files was generated with these default parameters and placed at [http://www.stsci.edu/science/starburst/PCStarburst99/.html](http://www.stsci.edu/science/starburst/PCStarburst99/.html). Users may want to compare these standard files to those produced on their PC. These standard files are not included in the zipped installation package since they would have added another ~50 MB of data.

A typical run of Starburst99 will allocate about 90 MB of virtual memory and utilize about 50% of the available CPU under no load. Windows normally will adjust the allocated CPU if other processes are running. In Table 4 we compare the run times of Starburst99 on three Windows systems using the default parameters. System 1 is a Micron Millennia desktop manufactured in 2003. It has a Pentium 4 3.2 GHz processor. Note that this is a single-core processor. The system has a total of 2 GB of random access memory (RAM) and an internal hard-drive with 300 GB of disk space. The operating system is XP Professional SP3. The run-time on this PC is 36 minutes. System 2 is a modern, 2008-built Dell Latitude D830 notebook. This system has an Intel dual-core 2.6 GHz processor, 4 GB of RAM and a 150 GB hard drive. It has Vista Ultimate SP1 installed. Starburst99 is running noticeably faster on this late-type notebook (purchased in summer 2008). The run-time is 18 minutes. Finally, System 3 is a Dell Optiplex 755 desktop with Vista Ultimate SP1 installed. It has an Intel dual-core 3.2 GHz processor and otherwise specifications similar to System 2. The somewhat faster processor leads to a shorter run-time of only 14 minutes. These run-times are at least an order of magnitude shorter than those on the STScI web server running the UNIX Solaris version of Starburst99 demonstrating the enormous benefit of the PC version for users. There are no significant numerical differences between the output generated on the PCs and that from the UNIX/Solaris environment.

## 6. Distribution

The PC version of Starburst99 is available as freeware to the astronomical community. The primary distribution channel is via the web. Interested users may download the package from the Starburst99 site at http://www.stsci.edu/science/starburst/PCStarburst99/.html. A limited number of CDs with the Starburst99v51.zip file is available. Users who cannot access the website may request their own copy of the CD. The zip file called Starburst99v51.zip contains three files:

- *setup.exe*: the program to launch the Windows installer;
- *Setup*.msi: the Windows installer for the Starburst99 application and all its required auxiliary files as well as the graphic user interface;
- *README*.txt: release notes and instructions for installing the application.

The Windows installer will install v5.1 of the Starburst99 software. A successful installation will show a root folder named Starburst99 which contains the five subfolders *auxil*, *help*, *lejeune*, *models*, and *tracks*, as well as four additional files:

- *galaxy.exe*: the executable for the Starburst99 population synthesis code;
- *galaxy.f*: the Fortran source code for galaxy.exe;

- *Starburst99.exe*: the program for launching the user interface;
- *Starburst99.ICO*: the desktop icon.

Once installed, user should not have to click on the executables manually but will rather use the graphic user interface to access all features of the Starburst99 package. Users are offered free registration of the software. Registered users will be notified of available updates, program errors, and new versions of the Starburst99 package. Limited email support is provided as well.

We are currently maintaining both the Windows and UNIX/Solaris versions of Starburst99. The two versions are kept up-to-date and consistent with each other. The currently existing Linux version of the code will no longer be supported and no new versions will be generated. It is planned to develop an Apple/Macintosh compatible version of the code in the future.

# Figures

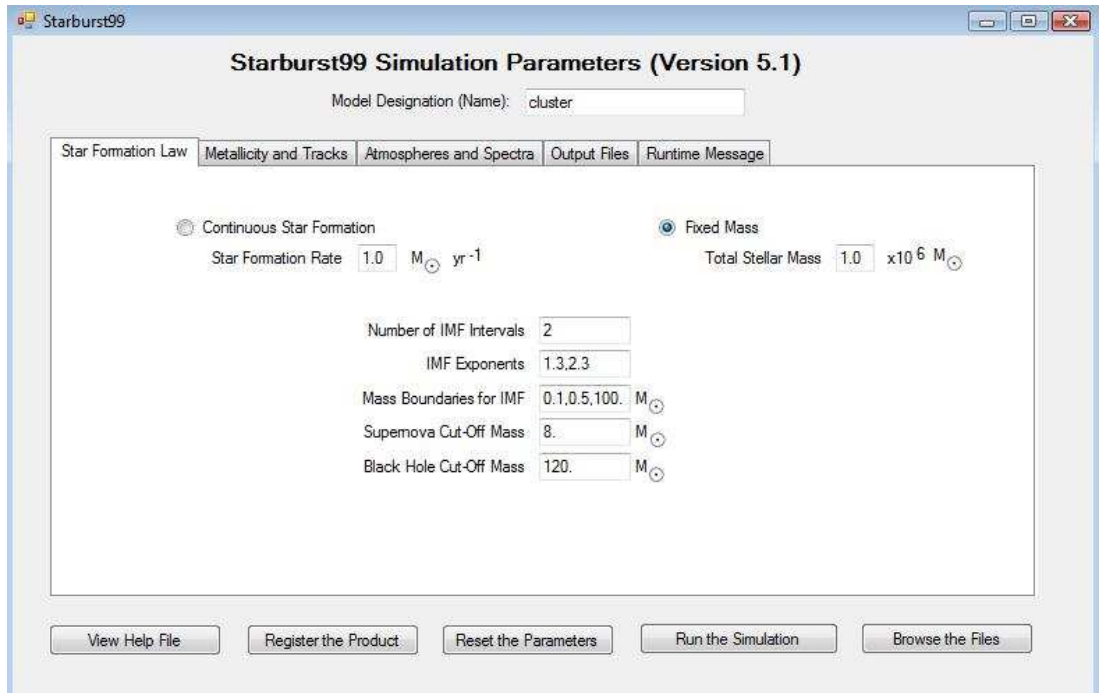

Figure 1. — User interface for entering parameters related to the star-formation law.

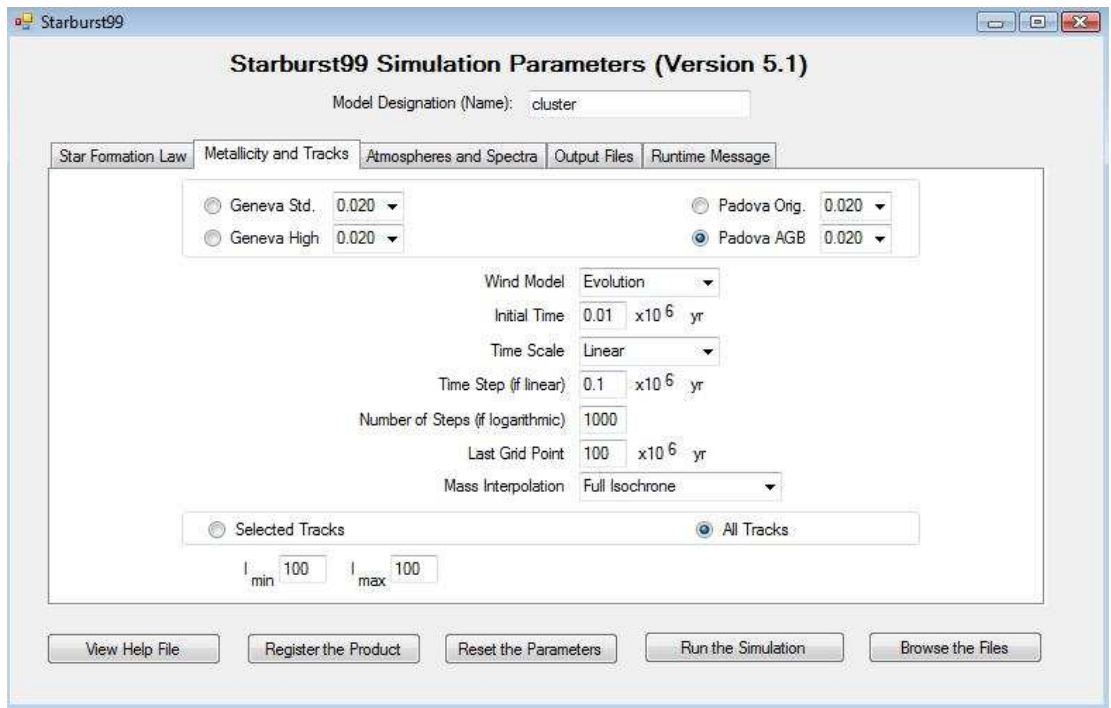

Figure 2. — User interface for entering parameters related to the stellar evolution models.

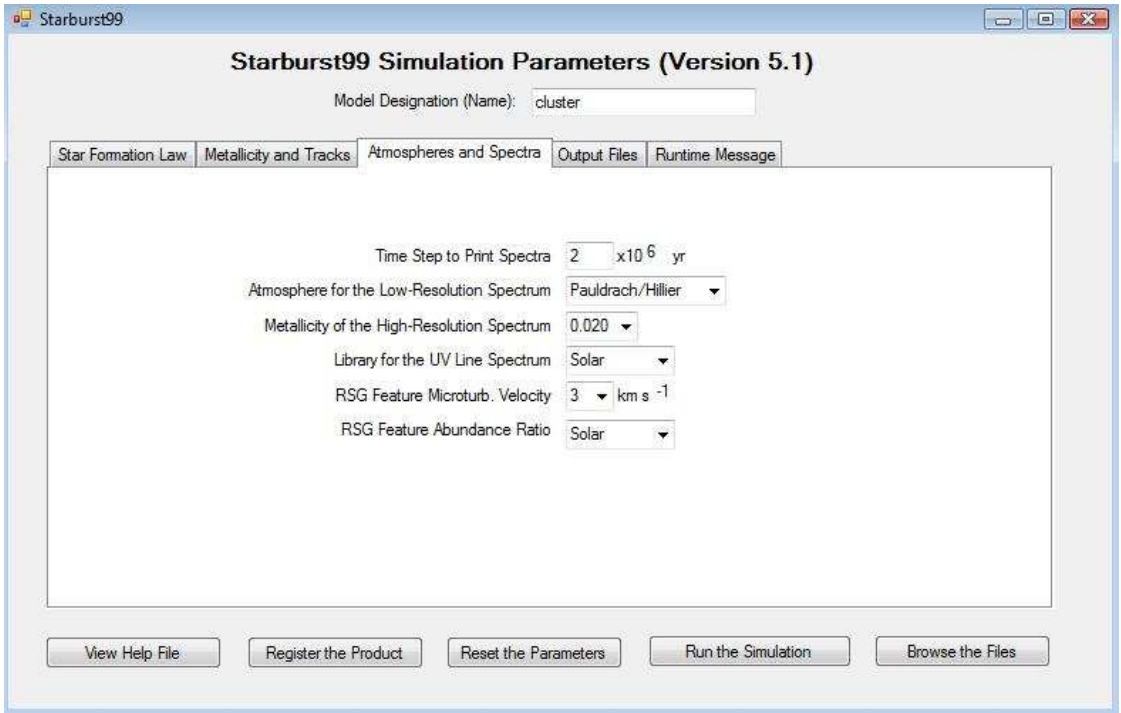

Figure 3. — User interface for entering parameters related to the stellar atmospheres.

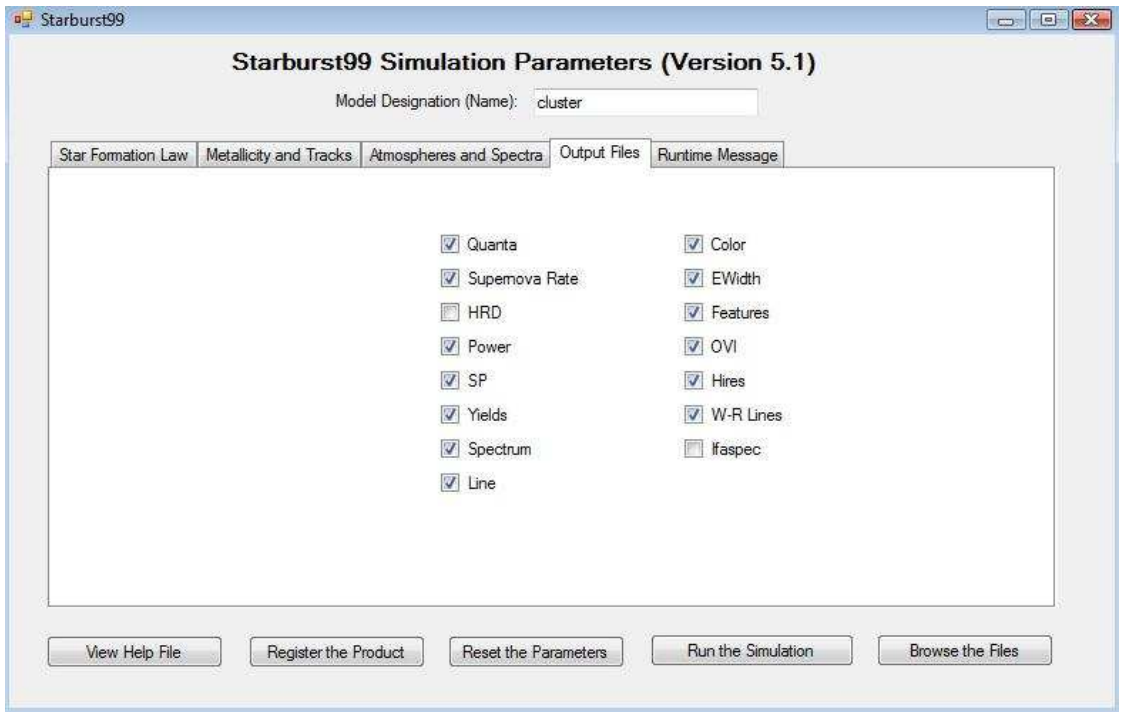

Figure 4. — Selection of the output files.

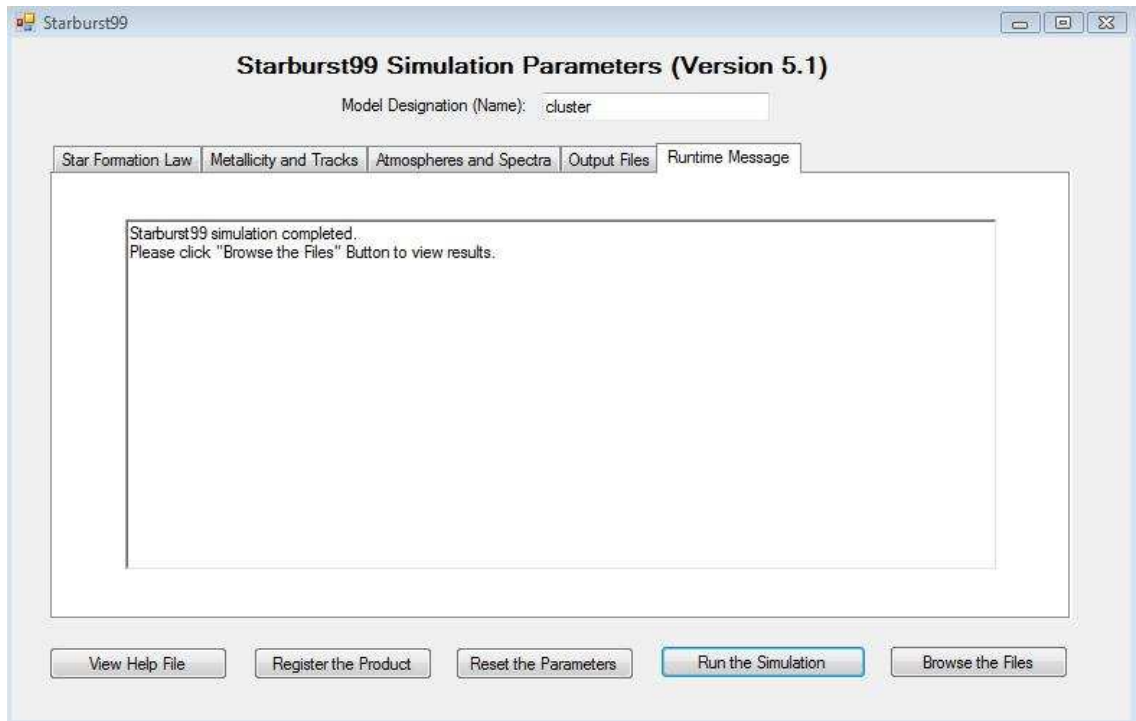

Figure 5. — Display of the Runtime Message

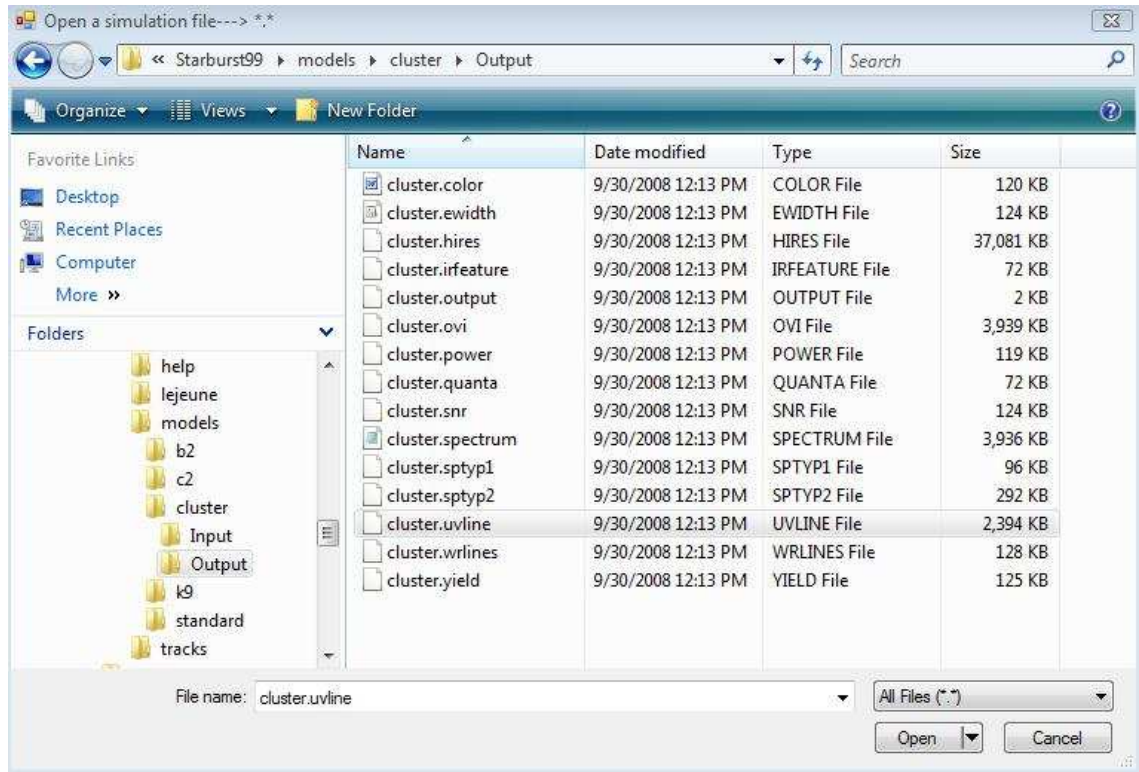

Figure 6. — Window for browsing and opening the output files.

# Tables

Table 1. Input files containing the stellar evolution models

| Name | Count | Placeholder Values | File Size | Description |
|---|---|---|---|---|
| modc*ZZZ*.dat | 4 | *ZZZ*=001,004,020,040 | 100 KB | Geneva, $1 \times \dot{M}$ |
| mode*ZZZ*.dat | 4 | *ZZZ*=001,004,020,040 | 100 KB | Geneva, $2 \times \dot{M}$ |
| modp*ZZZ*.dat | 5 | *ZZZ*=0004,004,008,020,050 | 310 KB | Padova, AGB |
| mods*ZZZ*.dat | 5 | *ZZZ*=0004,004,008,020,050 | 310 KB | Padova, no AGB |

Table 2. Input files containing the model atmospheres

| Name | Count | Placeholder Values | File Size | Description |
|---|---|---|---|---|
| allstarscont_*ZZZ*.txt | 4 | *ZZZ*=m10,m05,p00,p03 | 61,000 KB | high-res, continuum |
| allstarsflux_*ZZZ*.txt | 4 | *ZZZ*=m10,m05,p00,p03 | 61,000 KB | high-res, lines |
| allstarswave.txt | 1 | 9 | 150 KB | high-res, wave |
| cmfgen_WC_*ZZZZ*.dat | 5 | *ZZZ*=001,004,008,020,040 | 450 KB | UCL, WC stars |
| cmfgen_WN_*ZZZZ*.dat | 5 | *ZZZ*=001,004,008,020,040 | 450 KB | UCL, WN stars |
| ifa_cont_*ZZZ*.txt | 5 | *ZZZ*=m13,m07,m04,p00,p03 | 4,130 KB | IfA, continuum |
| ifa_line_*ZZZ*.txt | 5 | *ZZZ*=m13,m07,m04,p00,p03 | 4,130 KB | IfA, lines |
| ifa_wave.txt | 1 | 9 | 100 KB | IfA, wave |
| lcb97_*ZZZ*.flu | 5 | *ZZZ*=m13,m07,m04,p00,p03 | 6,100 KB | BaSeL atmospheres |
| wmbasic_OB_*ZZZZ*.dat | 5 | *ZZZ*=001,004,008,020,040 | 450 KB | UCL, WC stars |
| wr_beta1.fluxes | 1 | 9 | 8,700 KB | WR, $\beta$=1 |
| wr_beta2.fluxes | 1 | 9 | 6,400 KB | WR, $\beta$=2 |

Table 3. Input files containing the spectral libraries and look-up tables

| Name | Count | Placeholder Values | File Size | Description |
|---|---|---|---|---|
| fuse_*X*.dat | 2 | *X*=high,low | 6,100 KB | FUSE library |
| irfeatures.dat | 1 | 9 | 50 KB | near-IR lines |
| schkal.dat | 1 | 9 | 14 KB | spectral types |
| sp*X*.dat | 2 | *X*=" ",_low | 3700 KB | ultraviolet library |

Table 4.  Run time comparison

| Computer | OS | Processor | RAM | Hard drive | Run time |
|---|---|---|---|---|---|
| Micron Millennia 920i | XP Professional SP3 | Intel Pentium 4 3.2 GHz | 2 GB | 300 GB | 36 min |
| Dell Latitude D830 | Vista Ultimate SP1 | Intel Core 2 Duo 2.6 GHz | 4 GB | 150 GB | 18 min |
| Dell Optiplex 755 | Vista Ultimate SP1 | Intel Core 2 Duo 3.2 GHz | 4 GB | 150 GB | 14 min |